\documentclass[conference]{IEEEtran}
\IEEEoverridecommandlockouts
\usepackage{cite}
\usepackage{amsmath,amssymb,amsfonts}
\usepackage{algorithmic}
\usepackage{graphicx}
\usepackage{textcomp}
\usepackage{longtable}
\usepackage{array}
\usepackage{graphicx,color}
\usepackage{float}
\usepackage{enumerate}
\usepackage{makecell}
\usepackage{longtable}
\usepackage{graphicx}
\usepackage{lipsum}
\usepackage{float}
\usepackage{multicol}
\usepackage{multirow}
\usepackage{xcolor}
\def\BibTeX{{\rm B\kern-.05em{\sc i\kern-.025em b}\kern-.08em
    T\kern-.1667em\lower.7ex\hbox{E}\kern-.125emX}}
\begin{document}

\title{SDN-Based Dynamic Cybersecurity Framework of IEC-61850 Communications in Smart Grid\\
}

\author{\IEEEauthorblockN{Mansi Girdhar, Junho Hong, Wencong Su}
\IEEEauthorblockA{\textit{Department of Electrical and Computer Engineering} \\
\textit{University of Michigan-Dearborn}\\
Dearborn, USA \\
gmansi@umich.edu, jhwr@umich.edu, wencong@umich.edu} 
\and
\and
\IEEEauthorblockN{Akila Herath, Chen-Ching Liu}
\IEEEauthorblockA{\textit{Department of Electrical and Computer Engineering} \\
\textit{Virginia Tech}\\
Blacksburg, VA, USA \\
akilaasansana@vt.edu, ccliu@vt.edu}
}

\maketitle

\begin{abstract}
In recent years, critical infrastructure and power grids have experienced a series of cyber-attacks, leading to temporary, widespread blackouts of considerable magnitude. Since most substations are unmanned and have limited physical security protection, cyber breaches into power grid substations present a risk. 
Nowadays, the susceptibility of SDN architecture to cyber-attacks has exhibited a notable increase in recent years, as indicated by research findings. This suggests a growing concern regarding the potential for cybersecurity breaches within the SDN framework. In this paper, we propose a hybrid intrusion detection system (IDS)-integrated SDN architecture for detecting and preventing the injection of malicious IEC 61850-based generic object-oriented system event (GOOSE) messages in a digital substation. Additionally, this program locates the fault's location and, as a form of mitigation, disables a certain port. Furthermore, implementation examples are demonstrated and verified using a hardware-in-the-loop (HIL) testbed that mimics the functioning of a digital substation.
\end{abstract}


\section{Introduction}

With its dynamic, controllable, cost-effective, and flexible nature, software-defined networking (SDN) is an emergent paradigm that is suitable for the high-bandwidth, dynamic applications of today. Moreover, data traffic is gradually rising along with the number of intelligent electronic devices (IEDs), making intelligent communication networks more complex. Hence, OpenFlow-based SDN is an advanced approach to network architecture that separates the control and forwarding planes to increase network flexibility as opposed to the traditional network switches, where the data forwarding and data control planes are tightly coupled, and hence limits the scalability and functionality. SDN network controllers are able to ascertain the path taken by network packets over a network of switches using OpenFlow, the first industry-standard communications interface designed between the control and forwarding layers of an SDN architecture. It allows network switches and controllers to access a network switch or router's forwarding plane across the network. By allowing the control plane to manipulate packet forwarding, the SDN architecture makes it possible to directly program network control and abstract the underlying infrastructure for applications and network services. As a result, nowadays, SDN is applied in intelligent substations to enhance the functionality, accessibility, and adaptability of communication and to meet the demands of the intelligent substation communication network.

Despite being a promising technology, SDN faces security challenges that appear to be difficult to address. Kreutz's research has identified seven primary possible security vulnerabilities \cite{834546553}. Literature provides evidence that SDN allows for creating dynamic flow controls, network monitoring, network forensics, security policy modification, security service insertion, and basic network programming. However, attacks on control plane communications are one of the most pressing and challenging security issues that even exist today \cite{10226193}. The notion of logical control and forwarding function being separate increases the attack surface. Since SDN network management is stationed at the controller and controllers are vulnerable to various cyber-attacks, including fault injection and distributed denial-of-service (DDoS) attacks, it can potentially inflict significant harm. Thus, it is critical to detect such attacks in SDN for future network SDN deployments.

Efforts from power vendors and researchers are focused on enhancing reliability and security through technological advancements, intelligent frameworks, and leveraging data resources for smart substation development. A key priority is defending against substation cyber-attacks by implementing technologies such as intrusion detection systems (IDSs) to prevent malicious data intrusion.

\begin{table*}[hbt!]
\centering
\caption{Distribution of Cyber-Attacks and Proposed Mitigations against SDN Architecture.}
\label{tab:Attacks}
\begin{tabular}{|l|l|l|l|l|}
\hline
\multicolumn{5}{|c|}{\textbf{Distribution of Cyber-Attacks and Proposed Mitigations against SDN Architecture}} \\
\hline

\multicolumn{1}{|c|}{\textbf{Attack Category}} &
  \multicolumn{1}{c|}{\textbf{Authors}} &
  \multicolumn{1}{c|}{\textbf{Year}} &
  \multicolumn{1}{c|}{\textbf{Mitigation Strategy}} &
  \multicolumn{1}{c|}{\textbf{Reference}} 
  \\
 \hline
 \hline

  

 

\begin{tabular}[c]{@{}l@{}} Hijack \end{tabular}
 &  Li et al. &  2019 &
  \begin{tabular}[c]{@{}l@{}}A software-defined active synchronous detection
 is presented \\to protect networked microgrids. \end{tabular} & 
  \begin{tabular}[c]{@{}l@{}}\cite{67815328} \end{tabular} \\
 
  \hline



 \begin{tabular}[c]{@{}l@{}} DoS \end{tabular}
 &  Usman et al. &  2020 &
  \begin{tabular}[c]{@{}l@{}}A network IDS is used to protect virtualization servers from HTTP DoS\\ attacks.\end{tabular} & 
  \begin{tabular}[c]{@{}l@{}}\cite{9231699} \end{tabular} \\
 
  \hline

\begin{tabular}[c]{@{}l@{}} DoS \end{tabular}
 &  Grammatikis et al. &  2021 &
  \begin{tabular}[c]{@{}l@{}}SDN-microSENSE architecture, which introduces a set of cybersecurity\\ and privacy mechanisms.\\ \end{tabular} & 
  \begin{tabular}[c]{@{}l@{}}\cite{678452535} \end{tabular} \\
 
  \hline





\begin{tabular}[c]{@{}l@{}}FDI and Hijack \end{tabular}
  &  Guzmán et al. & 2022  &
  \begin{tabular}[c]{@{}l@{}}SDN-based host tracking service is introduced to prevent corruption of the \\inverter control parameters.\end{tabular} & 
  \begin{tabular}[c]{@{}l@{}}\cite{9923902} \end{tabular} \\

  \hline

  \begin{tabular}[c]{@{}l@{}}DDoS\end{tabular}                                      & Khedr et al.  & 2023                                                                     & \begin{tabular}[c]{@{}l@{}} A machine learning-based multi-layered FMDADM framework is used as \\mitigation.\end{tabular} & \begin{tabular}[c]{@{}l@{}} \cite{10078255}  \end{tabular}   \\

\hline

\hline
 \hline
 \end{tabular}
\end{table*}

Literature shows plethora of research works involving SDN-based mitigation schemes. Research in \cite{8944426} proposed an entropy-based solution for the mitigation of DDoS attacks on the control plane of the SDN architecture. The authors of \cite{9974724} provide a method for identifying and countering Slowloris, a low-level DDoS attack, in an SDN environment. In order to detect and mitigate packet injection threats, the concept in \cite{1007828655} creates PIEDefender, an effective, protocol-independent component installed on SDN controllers. A compilation of recent attacks on SDN architecture and their proposed mitigations is shown in Table~\ref{tab:Attacks}.

Numerous similar efforts are being made in this direction to build an IDS that is capable of identifying and preventing cyber-attacks in the control and data planes of SDN architecture. To the best of the authors' knowledge, no work has addressed attack localization in a substation, though. Further, generic object-oriented system event (GOOSE) messages are crucial for real-time communication in substation automation, but their vulnerability has marked them as a frequent target in cyber-attacks (as mentioned in many research works), emphasizing the critical need to address their security vulnerabilities. So, in this paper, a rule-based IDS \cite{6786500} integrated SDN  module is introduced that identifies abnormal GOOSE packets and accurately determines the fault device in the network, hence aiding in localization and response to cybersecurity threats.

The primary contributions of this paper are: (1) a rule-based IDS-integrated SDN framework to detect the malicious GOOSE communication, locate the attacked device, and disable the port associated with the fault, (2) validation of the proposed method on a hardware-in-the-loop (HIL)-testbed by implementing two case scenarios, that includes malicious GOOSE attack on station bus SDN switch and protection intelligent electronic device (PIED), respectively, (3) computation of time delays associated with or without the proposed methods.

The remainder of the paper is divided as follows: Section II presents the high-level overview of a digital substation. Section III outlines the proposed IDS-integrated SDN framework. Section IV and Section V focus on the existing HIL testbed and evaluate the performance of the proposed module by case study analysis, respectively. It also presents time delays for the system configuration- with and without the application of IDS-integrated SDN. Finally, Section VI concludes the paper along with recommendations for future work.

\section{High-Level Overview of Substation Automation System}
The IEC TC 57 published and recommended the IEC 61850 standard for substation applications. A substation automation system can be structured into three levels: the process, bay, and station levels. A user-interface system comprising a database, workstation, and engineering facilities is implemented at the station level. At the bay level, phasor measurement units (PMUs) and PIEDs are installed. Sensors, current transformers, potential transformers, circuit breakers (CBs), and merging units (MUs) are examples of process level devices. The communication protocols (IEC 61850 standard) for handling information between the devices are sampled values (SVs), GOOSE, and manufacturing message specification (MMS). The communication traffic at the process/station buses is controlled by network switching devices (e.g., traditional Ethernet and SDN switches). 

\vspace{-1.25 pt}
\section{Proposed Framework}
\vspace{-1.25 pt}
The primary objective of the proposed framework is to detect and mitigate cyber-attacks in a substation by reconfiguring the SDN switches. Fig. 1 illustrates the high level overview of the proposed IDS-integrated SDN framework that is used in this paper. It demonstrates the use of multiple components. These are: (1) a rule-based IDS \cite{6786500}, that detects malicious GOOSE packets in the transmission; (2) an SDN controller is shown that receives the control operation commands and configures the SDN switch to enable or disable the port. The framework, as illustrated in Fig. 1, is divided into four sections: (1) network traffic generation; (2) SDN switch/IED attack; (3) rule-based detection \cite{6786500} on IDS-integrated SDN; and (4) identification of malicious host behind the attack.

\begin{figure}[htb!]
\centering
\includegraphics[width= 0.21\textwidth, height = 1.20in]{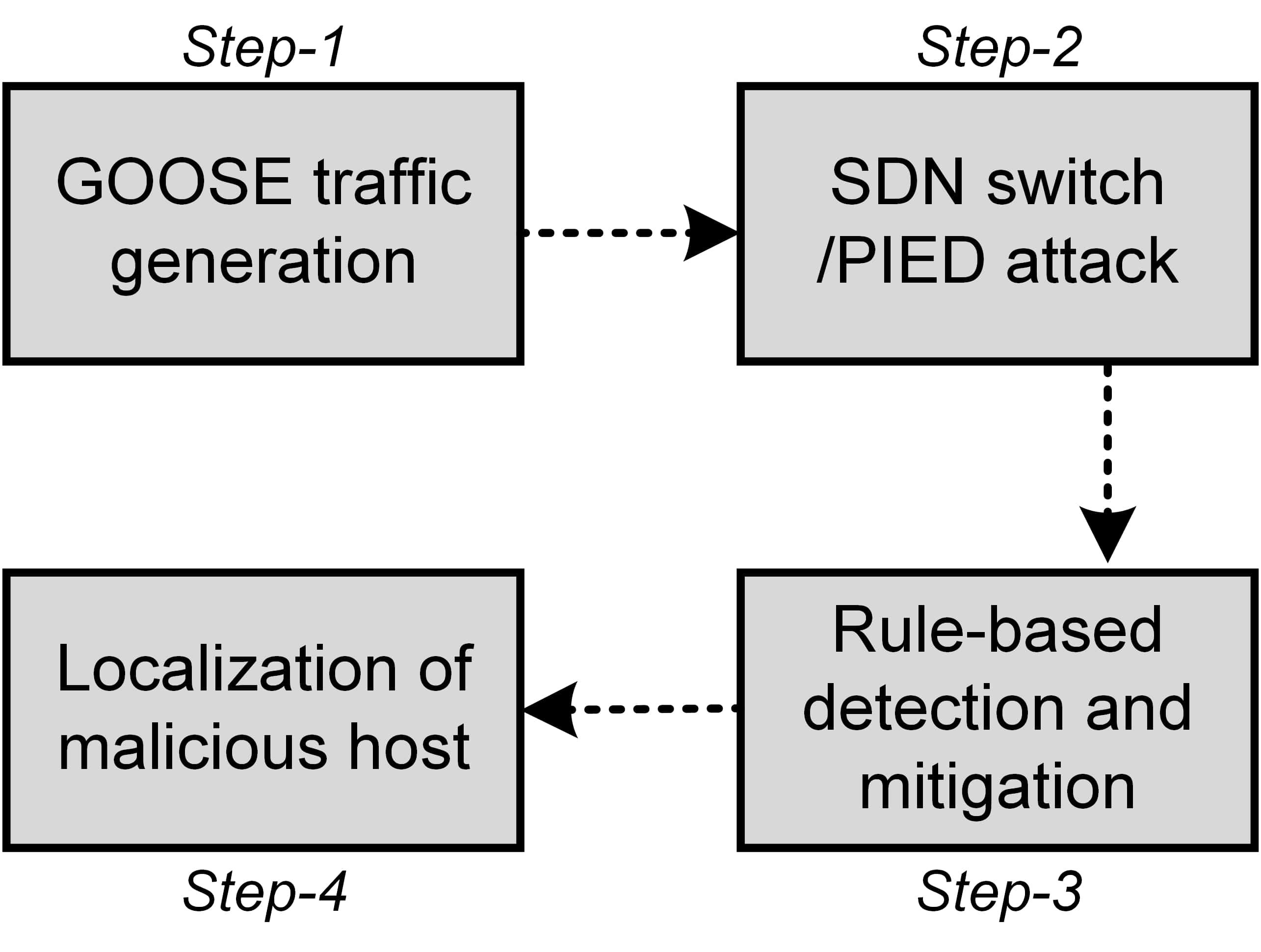}
\caption{Proposed framework.}
\label{fig:GM3}
\end{figure}
\vspace{-1.25 pt}

\section{Implementation Setup}

The HIL testbed utilizes the capabilities of a mininet emulator, in which Open vSwitch (OVS) switches are built and managed by an open-source SDN controller, named RYU on the Ubuntu platform. The OpenFlow protocol, which is exceptionally flexible for building custom networks, is used by mininets, which are supported by the LINUX platform. This paper uses a Python script to detect and prevent abnormal GOOSE packet injection attacks from compromised hosts (e.g., station bus SDN switch and PIED) that might be involved. The evaluation outcome demonstrates how the proposed IDS-integrated SDN can be applied to mitigate the impacts of an attack by measuring the time delays using the proposed IDS-integrated SDN device.

In a complex network system involving Omicron, MU, PIED, various SDN switches, and controllers, as shown in Fig. 2, a sequence of interactions occurs to manage and secure the network. Omicron sends analog three-phase currents and voltages, which are then received by the MU. The MU subsequently transmits SVs over the process bus to the PIED. In the event of an overcurrent fault detection, the PIED uses GOOSE messages to communicate with the Omicron, instructing it to trip the CBs.

\begin{figure}[htb!]
\centering
\includegraphics[width= 0.30\textwidth, height = 2.0in]{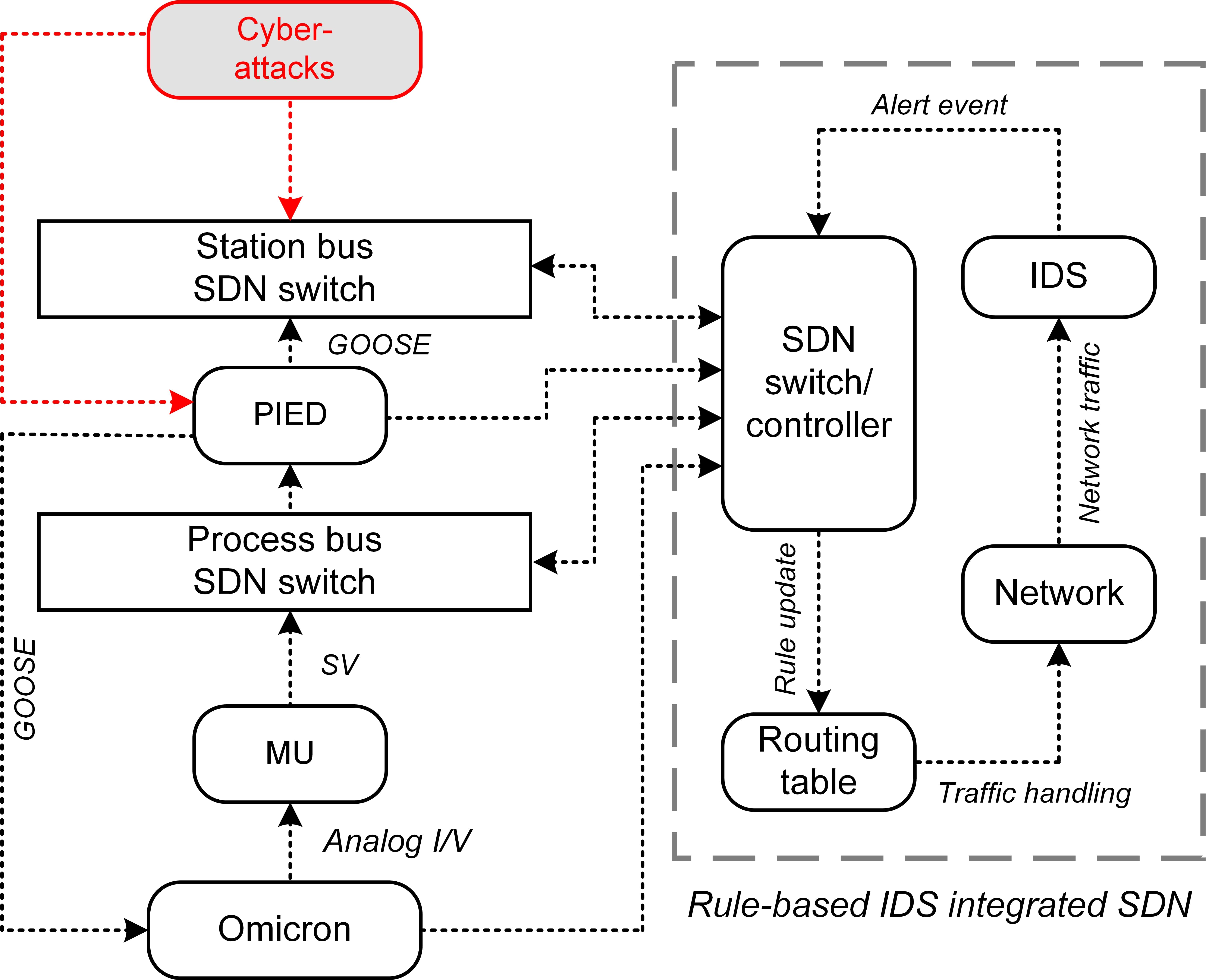}
\caption{HIL-testbed and basic system architecture with rule-based IDS-integrated SDN.}
\label{fig:GM1}
\end{figure}

The CB trip command, involving GOOSE messages, takes place over the station bus SDN switch. Moreover, the ports on station bus SDN switch, process bus SDN switch, PIED, and Omicron are responsible for transmitting vital information 
to an SDN switch/controller in a separate loop. This loop includes the SDN switch/controller, a rule-based IDS, the broader network, and the routing table.

The flow of information within IDS-integrated SDN follows a specific order. The SDN controller continuously collects and processes data from various sources, including MU, PIED, Omicron, and switches.  
The IDS, closely integrated into the SDN ecosystem, responds to detected anomalies and triggers an alert.  
Upon receiving an alert event from the IDS, the controller (IDS-integrated SDN) initiates a rule update process.
Finally, the proposed IDS monitors the network traffic for any anomalies or security threats based on the updated rules. After receiving the updated instructions, the switches now implement the changes in their flow tables. They make decisions based on the new rules, directing and forwarding traffic according to the altered configurations, which may involve prioritizing, redirecting, or blocking specific data flows based on the updated rules. 
By observing the changes in traffic patterns and the affected areas within the network in response to the fault event, the proposed SDN framework provides insights that can help identify the fault's specific location or segment. 
\vspace{-1.25 pt}
\section{Case Studies}
In the context of this paper, abnormal GOOSE packets are denoted as $P_{F}^{T}$, where ``F'' stands for ``From'' designates the source or origin of the malicious packets, and ``T'' stands for ``To'' designates the destination. The host devices (responsible for injecting abnormal GOOSE packets) are identified as station bus SDN switch ($S_{s}$) and PIED, collectively forming the set F and the IDS-integrated SDN ports ($IDS_{i}$) form the set T.
\vspace{-1.25 pt}
\begin{equation}
F = \left\{S_{s}, PIED \right\}.
\end{equation}
\vspace{-1.25 pt}
\begin{equation}
T = \left\{IDS_i \right\}, i = 1,2,...,8.
\end{equation}

So, in this work, the orchestrated actions of the SDN infrastructure—spanning from data analysis and rule updates to traffic handling adjustments—enable the network to effectively localize and identify the specific area or device impacted by a fault. 

\subsection{Attack Scenario 1: Abnormal GOOSE Injection Attack on Station Bus SDN Network Switch}

As illustrated in Fig. 3, 
the attacker might exploit vulnerabilities in the station bus SDN switch, gain unauthorized access, and inject abnormal packets without a physical connection at port 6. Alternatively, as an insider or through compromising a device, they might manipulate the station bus SDN switch \cite{9122482}. Inadequate mention of security measures, e.g., ingress blocking or susceptibility to advanced persistent threats, might further facilitate malicious packet injection. Hence, an IDS-integrated SDN is strategically placed to detect the source of abnormal malicious GOOSE packets. In this scenario, station bus SDN switch governed by flow tables, dynamically replicates packets when matching multiple flow table entries, allowing it to efficiently forward duplicated frames to multiple destination ports simultaneously. The sequence of steps and the IDS's roles (determining the abnormal packet's origin) are detailed below:
\begin{figure}[htb!]
\centering
\includegraphics[width= 0.30\textwidth, height = 2.00in]{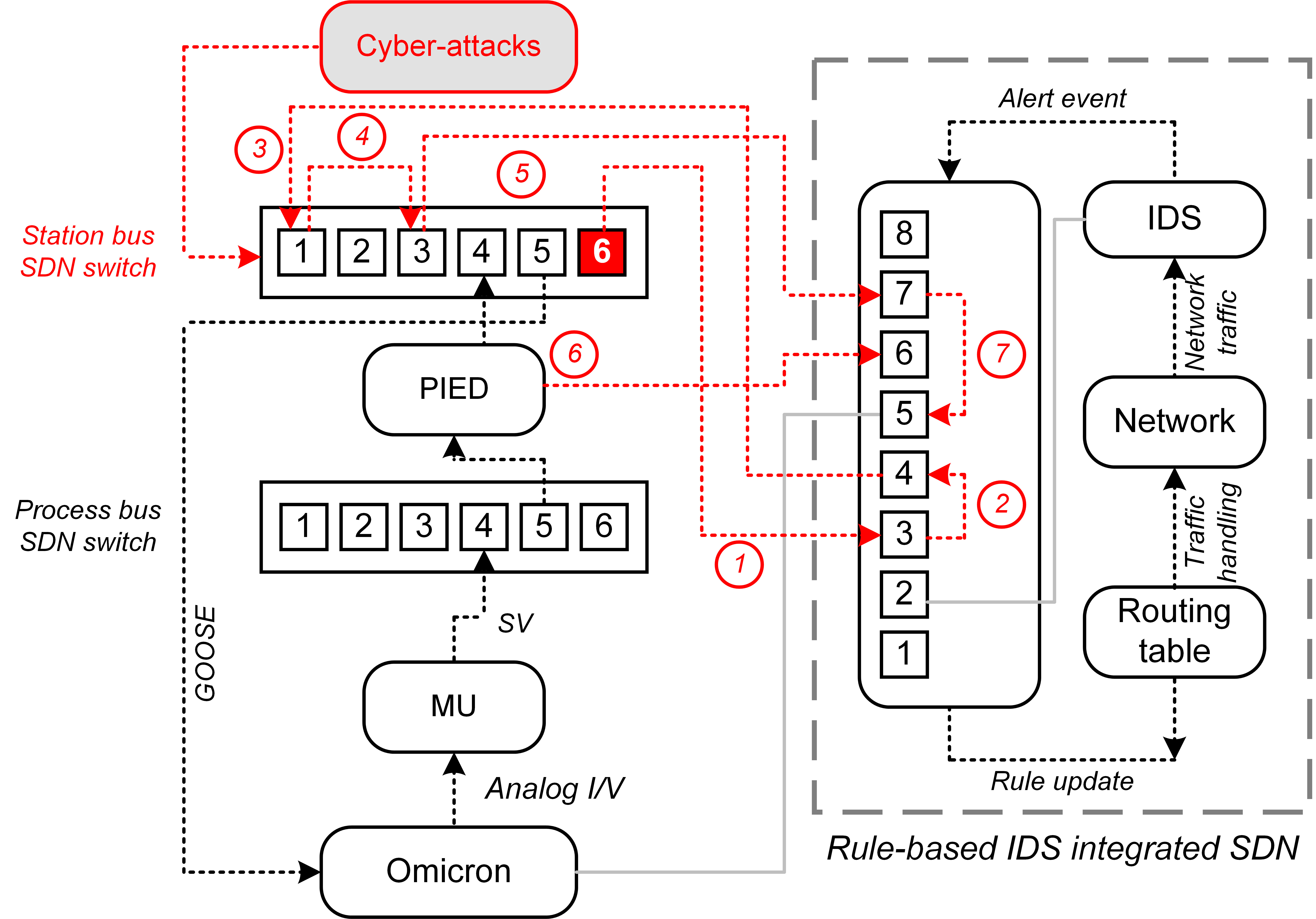}
\caption{Attack Scenario 1: abnormal GOOSE injection attack on station bus SDN network switch.}
\label{fig:CS1_1}
\end{figure}

\begin{enumerate}
    \item {Abnormal Packet Generation from Station Bus SDN Switch (Port 6):} Attacker generates abnormal GOOSE packets directly from the station bus SDN switch, utilizing port 6 for this purpose.

\item{IDS Inspection at Port 3:} Once the proposed IDS-integrated SDN receives the GOOSE packet, it will duplicate the packets to port 4 internally for further analysis and start the packet inspection. The IDS processing block is responsible for inspecting the packet, applying detection algorithms, and determining whether the packet exhibits characteristics of a malicious GOOSE packet.

\item{Forwarding to Station Bus SDN Switch (Port 1):} IDS sends the packet to the station bus SDN switch, from port 4 (IDS) to port 1 (station bus).

\item{Station Bus SDN Switch Duplication to Port 3:} Station bus SDN switch receives the packet and forwards it to IDS, from port 3 (station bus) to port 7 (IDS).

\item{Packet Duplication to IDS at Port 7:} SDN switch then forwards the packet to the IDS at port 7 for further process, allowing the IDS to monitor both port 3 and port 7 concurrently.

\item{Normal GOOSE Packet Generation from PIED to IDS at Port 6:} PIED also sends GOOSE packets to IDS-integrated SDN to port 6 (IDS). Hence, IDS monitors port 3, port 6, and port 7 simultaneously.

\item{Forwarding to Omicron at Port 5:} The IDS (equipped to monitor port 3, port 6, and port 7) uses its inspection capabilities to identify the source of the abnormal packet. In this context, it can identify the origin of the abnormal packets (station bus or PIED). Then, it forwards the packet to port 5, facilitating its transmission to the Omicron for further analysis or action.
    
\end{enumerate}
To summarize, the IDS-integrated SDN monitors port 3, port 6, and port 7.
 Since it receives the abnormal GOOSE packets from the station bus SDN switch through port 7 ($P_{S_{s}}^{IDS_{7}}$) and port 3 ($P_{S_{s}}^{IDS_{3}}$). Hence, it localizes the station bus SDN switch as the malicious host. After identifying the compromised station bus SDN switch, IDS will disable all communication ports with the compromised devices and then only enable IDS ports 5 and 6 for normal operation. 

\subsection{Attack Scenario 2: Abnormal GOOSE Cyber-Attack on PIED}
As shown in Fig. 4, cyber-attackers successfully compromised the PIED and its communication. The subsequent actions are delineated in the following steps:\\

\begin{figure}[htb!]
\centering
\includegraphics[width= 0.30\textwidth, height = 2.00in]{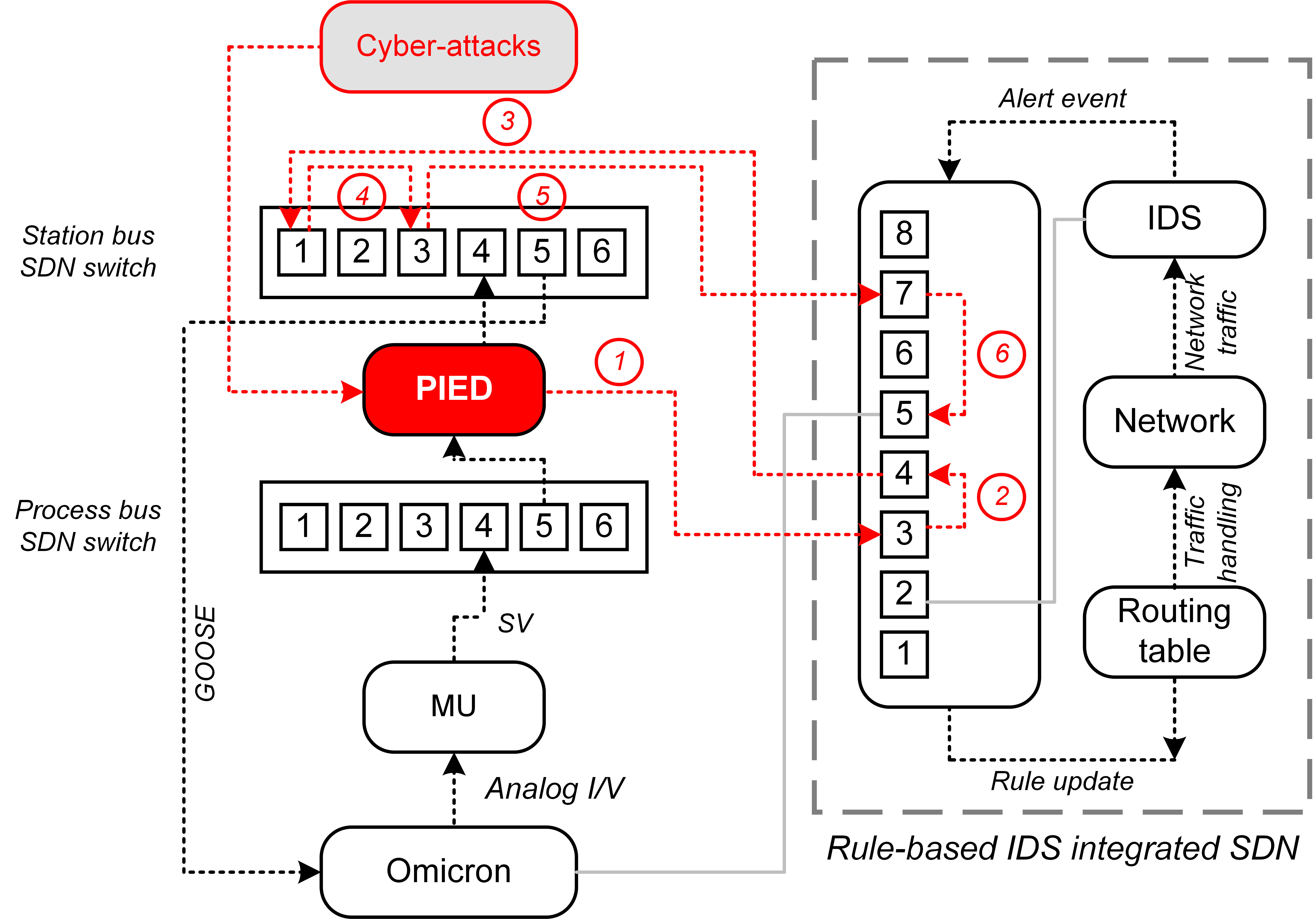}
\caption{Attack scenario 2: abnormal GOOSE cyber-attack on PIED.}
\label{fig:CS2}
\end{figure}

\begin{enumerate}
    \item {Generation of Abnormal Malicious GOOSE Packets:} The cyber-attacker initiates malicious activity by generating abnormal GOOSE packets from the PIED.

\item{IDS Inspection at Port 3:} IDS receives the GOOSE packet, actively inspects the incoming packet at port 3, and subsequently forwards it to port 4.

\item{Forwarding to Station Bus SDN Switch (Port 1):} IDS sends the packet to the station bus SDN switch, from port 4 (IDS) to port 1 (station bus).

\item{Station Bus SDN Switch Forwarding to Port 3:} Station bus SDN switch receives the packet and forwards it from port 1 (station bus) to port 3 (station bus).

\item{Packet Redirection to IDS at Port 7:} SDN switch then forwards the packet to the IDS at port 7, thereby enabling the IDS to monitor both port 3 and port 7. This configuration allows the IDS to distinguish whether the abnormal GOOSE packet originates from the PIED (monitored at port 3) or the station bus (monitored at port 7).

\item{Identification and Forwarding to Omicron at Port 5:} Upon differentiation, the IDS identifies the packet's source and forwards it to port 5 (IDS), subsequently sending it to the Omicron.

\end{enumerate}
In summary, the IDS-integrated SDN 
monitors both port 3 and port 7.
Since, it receives abnormal GOOSE packet from PIED through port 3 ($P_{PIED}^{IDS_{3}}$). Hence, it localizes PIED as the malicious host. After identifying the compromised PIED, IDS will disable all communication ports with the compromised devices (station bus port 4 and process bus port 5) and then successfully isolate the compromised PIED.

 \subsection{Computation of Time Delays}
Time delays related to communication are calculated for both scenarios—having the IDS-integrated SDN module within the framework and without it. 
\begin{figure}[htb!]
\centering
\includegraphics[width= 0.40\textwidth, height = 1.75in]{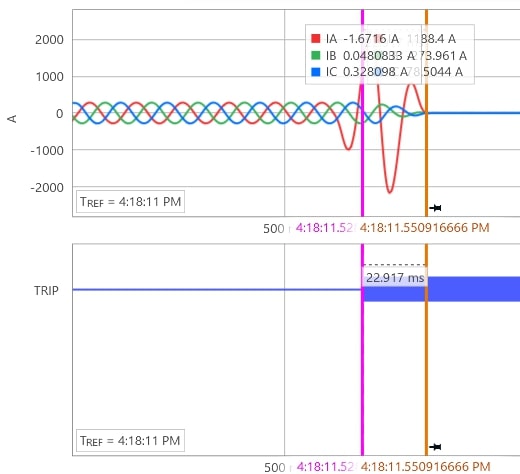}
\caption{Time delay in a normal scenario without the proposed IDS-integrated SDN module.}
\label{fig:Time_Delay04}
\end{figure}

In the typical situation, where the proposed module is not in use, the average time is measured between the tripping and closing of the CB and is determined to be approximately 23ms (ignoring mechanical delay of CB). Fig. 5 clearly illustrates the time delay ($T_{Total}$) associated with it. The total time delay
includes internal delays in the MU ($T_{MU}$), SV communication delay ($T_{SV}$), process bus SDN switch routing delays ($T_{S_p}$), delays in PIED ($T_{PIED}$) to calculate the overcurrent fault protection algorithm, station bus SDN switch ($T_{S_s}$), GOOSE communication delay ($T_{GS}$), and internal Omicron delays ($T_{OC}$).
In an alternative scenario following the integration of the proposed IDS-integrated SDN module into the system, it has been noted that the overall time delay ($T_{Total}$), as shown in Fig. 6, remains largely consistent, with the maximum value of approximately 27ms. This is primarily due to the minimal delays introduced into the network due to SDN routing and packet forwarding in the IDS-integrated SDN module, as indicated by (\ref{eqn2}).
Consequently, it does not exert extreme influence on the timing delays (less than a quarter cycle, 60Hz).
\begin{equation}
\label{eqn2}
\begin{split}
 \mathrm{T}_{Total}= \mathrm{T}_{MU} + \mathrm{T}_{SV} + \mathrm{T}_{S_{p}} + \mathrm{T}_{PIED} \\ + \mathrm{T}_{S_{s}} + \mathrm{T}_{GS} + \mathrm{T}_{OC} + (\mathrm{T}_{IDS}).
 \end{split}
\end{equation}

\begin{figure}[htb!]
\centering
\includegraphics[width= 0.40\textwidth, height = 1.75in]{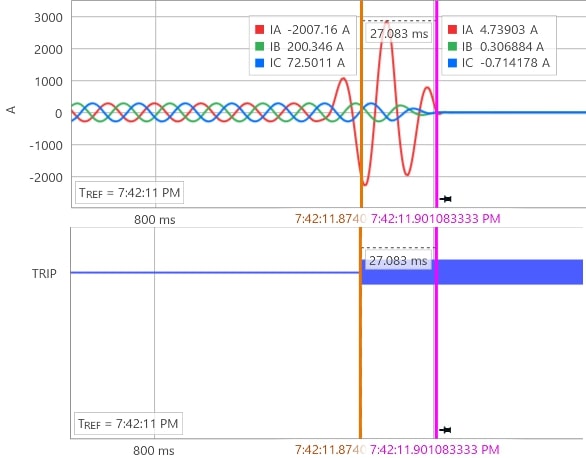}
\caption{Time delay with proposed IDS-integrated SDN module.}
\label{fig:Time_Delay03}
\end{figure}

\section{Conclusion and Future Work}
This paper introduces an IDS-integrated SDN architecture that aims to detect the injection of malicious IEC 61850-based GOOSE messages and locate faulty devices. Also, the IDS-integrated SDN framework is flexible, ensuring it can be applied to a broad spectrum of substation network configurations. The key focus lies in the fact that the introduced delay, stemming from the integration of the proposed IDS-integrated SDN architecture, is a result of the additional routing and processing carried out by the SDN-based IDS. The highest total delay to use the proposed IDS-integrated SDN framework is when a multiple routing method is chosen for the packet inspection algorithm, and this can be used for applications with a total of 4ms delay. Moreover, a similar IDS-integrated SDN architecture could be designed to determine if the malicious SV packets are coming from process bus SDN switches or MUs. As for future prospects, continuous refinement and optimization of the proposed IDS-integrated SDN architecture to enhance its accuracy and efficiency in detecting and preventing cyber threats could be considered. This includes fine-tuning the system's capabilities and response mechanisms. Further, as cyber threats continue to evolve, it's crucial to ensure that the system remains adaptable to new attack vectors and techniques. This might involve constant updates, machine learning, or AI-based mechanisms for threat detection. While GOOSE is the primary focus of this research, future work may explore the security implications of SV and MMS protocols that can be implemented with the similar framework. 

\section{Acknowledgement}
This research was supported in part by the Director, Cybersecurity, Energy Security, and Emergency Response, Cybersecurity for Energy Delivery Systems program, of the U.S. Department of Energy, under contract DE-CR0000021. Any opinions, findings, conclusions, or recommendations expressed in this material are those of the authors and do not necessarily reflect those of the sponsors of this work.

\bibliographystyle{IEEEtran}


\end{document}